\documentclass[aps, reprint, onecolumn, showpacs, superscriptaddress,
prl, longbibliography]{revtex4-2}
\usepackage {graphicx}%
\usepackage[caption=false, singlelinecheck=false]{subfig}
\usepackage{array}
\usepackage{wrapfig}
\usepackage{color}
\usepackage{latexsym}
\usepackage{bm}
\usepackage{courier}
\usepackage{gensymb}
\usepackage{import}

\usepackage[latin1]{inputenc}
\usepackage[T1]{fontenc}
\usepackage[normalem]{ulem}
\usepackage{verbatim}
\usepackage{amssymb}
\usepackage{amsmath}
\usepackage{wasysym}
\usepackage{epstopdf}

\renewcommand*{\v}[1]{\boldsymbol{#1}}
\newcommand*{\grad}{\v \nabla}

\begin{document}
\title{Self-similar and disordered front propagation in a radial Hele-Shaw channel with time-varying cell depth}
\author{C. Vaquero-Stainer}
\affiliation{Manchester Centre for Nonlinear Dynamics and
School of Physics and Astronomy, University of Manchester, Oxford Road, Manchester
M13 9PL, United Kingdom}
\author{M. Heil}
\affiliation{Manchester Centre for Nonlinear Dynamics and
School of Mathematics, University of Manchester, Oxford Road, Manchester
M13 9PL, United Kingdom}
\author{A. Juel}
\affiliation{Manchester Centre for Nonlinear Dynamics and
School of Physics and Astronomy, University of Manchester, Oxford Road, Manchester
M13 9PL, United Kingdom}
\author{D. Pihler-Puzovi\'{c}}
\affiliation{Manchester Centre for Nonlinear Dynamics and
School of Physics and Astronomy, University of Manchester, Oxford Road, Manchester
M13 9PL, United Kingdom}



\begin{abstract}
The displacement of a viscous fluid by an air bubble in the narrow gap
between two parallel plates can readily drive complex interfacial 
pattern formation known as viscous fingering. We focus on a 
modified system suggested recently by \cite{Zheng15}, in which
the onset of the fingering instability is delayed by introducing a 
time-dependent (power-law) plate separation.
We perform a complete linear stability analysis 
of a depth-averaged theoretical model to show that the plate separation 
delays the onset of non-axisymmetric instabilities, in
qualitative agreement with the predictions obtained from a simplified
analysis by \cite{Zheng15}.
We then employ direct numerical simulations to show that
in the parameter regime where the axisymmetrically expanding air bubble is 
unstable to non-axisymmetric perturbations, the interface can evolve 
in a self-similar fashion such that the interface shape at a given  
time is simply a rescaled version of the shape at an earlier time.
These novel, self-similar solutions are linearly stable but they only 
develop if the initially circular interface is subjected to unimodal
perturbations. Conversely, the application of non-unimodal
perturbations (e.g. via the superposition of multiple linearly
unstable modes) leads to the development of complex, constantly 
evolving finger patterns similar to those that are typically observed in 
constant-width Hele-Shaw cells.
\end{abstract}

\maketitle

\section{Introduction}
An expanding air bubble that displaces a viscous fluid within the
narrow gap between two parallel plates (a Hele-Shaw cell) is unstable
to non-axisymmetric perturbations beyond a critical value of the
capillary number $Ca$, the ratio of viscous to surface tension
forces~\cite{Saffman58, Paterson81, Chen89, Couder00}. In this radial
geometry, the unstable interface typically deforms into a set of
growing fingers which evolve continuously in time through a sequence
of tip-splitting events followed by competition between the 
newly-formed fingers. This non-linear evolution is an archetype for
front-propagating, pattern forming phenomena \cite{Swinney2003}.

Long-standing interest in viscous fingering stems from its
similarities with a range of other front propagation phenomena, such
as the growth of bacterial colonies~\cite{BenJacob92} and the
solidification instabilities during crystal
growth~\cite{Mullins64}. In hydraulic fracture used for oil recovery,
viscous fingering is promoted to ensure heterogeneous placement of
particles (proppants) into fractures, thereby increasing their
hydraulic conductivity, but it is also actively suppressed to avoid
early breakthrough of water into adjacent production
wellbores~\cite{Seright06, Osiptsov17}. This has stimulated a recent
resurgence in research effort to delay the onset of viscous fingering,
e.g. by manipulating the geometry of the cell, either 
actively~\cite{AlHousseiny12,  Bongrand18}, or passively by using 
compliant cells~\cite{PihlerPuzovic12,
  PihlerPuzovic18, Juel18}; by controlling the injection rate~\cite{Dias12}; or
by tuning the viscosity ratio of miscible fluid
pairs~\cite{Bischofberger14}.

In this paper, we consider fingering in a radial Hele-Shaw cell in which
the width of the gap between the parallel bounding plates increases 
as a function of time. This effect is stabilising since it reduces the
rate at which the axisymmetric bubble expands; in fact, if done 
sufficiently rapidly,
the plate separation can bring the expansion of the bubble to a halt or
even cause the bubble to contract.
In a recent study, \citet{Zheng15} employed 
a simplified analysis, based on \citeauthor{Paterson81}'s classical 
results for the system with constant gap width \cite{Paterson81}, to
examine the case when the gap width increases according to the power law
$b^*(t^*) = b_1^* \ {t^*}^{1/7}$, where $b^*$ is the distance between the 
plates and $t^*$ is time. They not only confirmed that viscous fingering can 
be suppressed if the plates are separated sufficiently rapidly (i.e. 
for sufficiently large values of the constant $b_1^*$), but also
predicted that the wavenumber $k_{\rm max}$ of the most rapidly-growing 
small-amplitude
perturbation to the axisymmetric bubble is independent of the bubble's 
radius. Furthermore, they performed experiments which showed that in 
the parameter range for which the axisymmetric bubble is
unstable, a small number of finite-amplitude non-splitting fingers tended to 
develop and that the number of these fingers
was remarkably close to $k_{\rm max}$. 

The observation of non-splitting fingers is unusual because it
contrasts with the typical sequence of tip-splitting and finger
competition observed in experiments without plate
lifting~\cite{Chen89, Couder00}.  The findings of \citet{Zheng15}
suggest that, following some initial transients, the finger growth may
become self-similar in the sense that the finger shape at a given time
is simply a rescaled version of the shape at an earlier time. This is
analogous to proportionate growth observed in biological systems,
e.g. during growth of a mammal whose body parts grow at nearly the
same rate and thus in direct proportion to each
other~\cite{Sadhu12}. In radial Hele-Shaw cells without lifting,
proportionate growth of fingering (or self-similar fingering) has been
observed in some experiments with miscible
fluids~\cite{Bischofberger14}.  Self-similar growth of the interface
between two immiscible fluids
has previously been observed~\cite{Thome89, Lajeunesse00} and
predicted~\cite{Combescot91, BenAmar91a, BenAmar91b} in a related 
setup where the cell consists of a disk sector.
This geometry promotes
the formation of a single finger which is symmetric about the sector
centreline and which does not split for sufficiently small values of
$Ca$ or sector angles. Other, more complex but well-defined,
reproducible finger shapes were also obtained by imposing selected
initial perturbations. Moreover, \citet{Li09} have computed
self-similarly evolving fingering patterns in a radial geometry by
appropriately varying the air injection rate, although
the number of fingers observed in that study differed from the most 
unstable wavenumber predicted by linear stability analysis.

In this paper we revisit the scenario studied by \citet{Zheng15}
and perform a full linear stability analysis to confirm that 
$k_{\rm max}$ is indeed independent of the bubble radius. 
We then perform numerical simulations of the system's nonlinear
evolution following the onset of the linear instability. We demonstrate that
self-similarly evolving fingering patterns can be realised with the
lifting law introduced by \citet{Zheng15}, but only when
the initially circular interface is perturbed with a
single, linearly unstable mode. Unimodal perturbations
with wavenumber $k_{\rm max}$ lead to the development of 
$k_{\rm max}$ distinct non-splitting, self-similar finite-amplitude fingers.
If the interface is subjected to unimodal perturbations with 
other wavenumbers the perturbed
interface typically undergoes a transient phase during which
fingers split before they approach a self-similar regime.
Therefore, the number of self-similar fingers that
emerge from the instability does not necessarily coincide with the
most unstable wavenumber predicted by the linear stability analysis. 
We also find that if non-unimodal, random initial perturbations are
introduced, such as those occurring in a typical experiment, the
interface does not evolve towards a self-similar solution, in contrast
with the scenario suggested by the experiments of
\citet{Zheng15}. Instead, the fingering pattern evolves continuously
as the interface advances across the cell through a succession of
tip-splitting and finger competition events similarly to the patterns
in a radial cell with fixed parallel plates. 

\section{Mathematical model}\label{model}
\begin{figure}
   \includegraphics{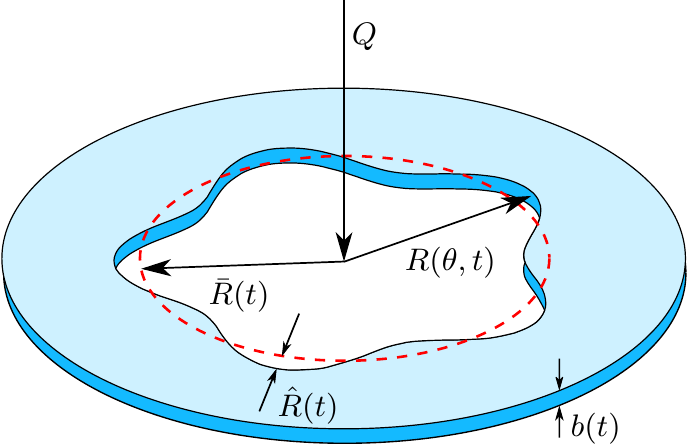}
  \caption{Schematic diagram of a perturbed interface in the radial
    Hele-Shaw cell with time-varying gap width.}\label{fig:setup}
\end{figure}
We describe the motion of the wetting
viscous fluid occupying the narrow gap between the two bounding plates
using the lubrication approximation. Using 
horizontal polar coordinates ($r^*, \theta$), centred at the
injection point (see figure~\ref{fig:setup}), the equations for
the vertically averaged velocity $\v u^*(r^*,\theta,t^*)$ and the
fluid pressure $p^*$ are then 
\begin{equation}\label{lubrication}
  \v u^* = -\frac{{b^*}^2}{12\mu}\grad p^*, \qquad \frac{\text{d}
    b^*}{\text{d} t^*} = -\grad\cdot (b^*\v u^*) =
  \frac{{b^*}^3}{12\mu}\grad^2p^* \qquad \text{ in } r^* >
  R^*(\theta,t^*),
\end{equation}
where $\mu$ is the liquid viscosity, $R^*(\theta,t^*)$ is the bubble
radius, and $b^*(t^*)$ is the (spatially constant) gap width. 
Throughout this paper
we employ asterisks to distinguish dimensional quantities from their
non-dimensional equivalents. In the
region $r^* < R^*(\theta,t^*)$, the gas bubble has a uniform pressure
$p_g^*(t^*)$. We neglect the effects of the thin films of liquid that
are left behind the advancing bubble tip, and ignore viscous normal
stresses at the interface. The kinematic and dynamic boundary
conditions at $r^* = R^*(\theta, t^*)$ then become
\begin{align}\label{bc}
 \frac{\partial \v R^*}{\partial t^*} \cdot \v n & = \v u^* \cdot \v
 n,\\ p^*(R^*, \theta, t^*) & = p^*_g(t^*)-\gamma
 (\kappa^*_{\perp}+\kappa^*_{\parallel})=p^*_g(t^*) - \gamma
 \left(\frac{{R^*}^2+2(\frac{\partial R^*}{\partial
     \theta})^2-R^*\frac{\partial^2 R^*}{\partial
     \theta^2}}{({R^*}^2+(\frac{\partial R^*}{\partial
     \theta})^2)^\frac{3}{2}} + \frac{2}{b^*}\right),
\end{align}
where $\v n$ is the unit normal to the interface and $\gamma$ is the
surface tension. The mean curvature of the interface is approximated
as the sum of the in-plane and transverse curvatures,
$\kappa^*_{\parallel}$ and $\kappa^*_{\perp}$, respectively. At the
outer boundary of the Hele-Shaw cell we impose $p^*(R^*_{\rm outer},
\theta, t^*)=0$. Finally, for a constant injection flux $Q$, the
volume of gas in the bubble is given by
\begin{equation}\label{mass_conservation}
Q t^* = \frac{b^*(t^*)}{2}\int_0^{2\pi}{R^*}^2(\theta, t^*)d\theta.
\end{equation}
We study the system's evolution starting from $t^*=t_0^*$ when the 
cell walls are separated by $b^*_0$ and the bubble has an initial
radius $R^* = R^*_{\rm init}$.

In the following analysis, we non-dimensionalize in-plane lengths with
$R^*_{\rm outer}$, the gap width with $b^*_0$, time with $2\pi
R^{*2}_{\rm outer}b^*_0/Q$ and pressures with $6\mu Q/\pi
b^{*3}_0$. Then the problem is governed by three non-dimensional
parameters: the capillary number $Ca = \mu Q/(2\pi \gamma R^{*}_{\rm
outer} b^*_0)$, the cell aspect ratio $\mathcal{A}=b^*_0/R^*_{\rm
outer}$ and the initial radius of the bubble $R_{\rm init} = 
R^*_{\rm init}/R^*_{\rm outer}$. At the initial time, when $t = t_0 = 
R_{\rm init}^2/2 $, we then have $b(t_0)=1$. When imposing
 \citeauthor{Zheng15}'s \cite{Zheng15} power-law 
behaviour for the plate separation
we have $b(t)=b_1 \ t^{1/7}$, where 
$b_1 = b_1^{*} \ (2\pi R^{*2}_{\rm outer}/(b_0^{*6} Q))^{1/7}
= t_0^{-1/7} = (R^2_{\rm init}/2)^{-1/7}$.

\section{Axisymmetric solutions and linear stability analysis}\label{LSA}
Equations (\ref{lubrication})-(\ref{mass_conservation}) have an
axisymmetric solution for which the bubble radius is given by
\begin{equation}
\label{axisym_radius}
\bar{R}(t)=\left(\frac{2t}{b(t)}\right)^{1/2}
\end{equation}
while the pressures in the viscous fluid and the gas bubble are
\begin{equation}
\bar{p}(r, t) = \frac{1}{b^3}\frac{\text{d} b}{\text{d}
  t}\frac{r^2-1}{4}-\frac{1}{b^3}\log{r},
\end{equation} 
and
\begin{equation}
p_g(t) = \frac{1}{b^3}\frac{\text{d} b}{\text{d}
  t}\frac{\bar{R}^2-1}{4}-\frac{1}{b^3}\log{\bar{R}}+
\frac{\mathcal{A}}{12Ca}\left(\frac{\mathcal{A}}{\bar{R}}+\frac{2}{b}\right),
\end{equation} 
respectively. To assess the stability of this solution to
non-axisymmetric perturbations that change the shape of the
interface, we assume $R(\theta, t) = \bar{R}(t) +
\varepsilon\hat{R}_k(t)\sin (k\theta)$, where $\varepsilon \ll 1$ is
the amplitude and $k>1$ is the integer wavenumber of the perturbation. 
A straightforward linear stability analysis then shows that the
instantaneous growth rate $\lambda_k$ of the small-amplitude 
perturbation with wavenumber $k$ is given by
\begin{equation}
\label{growthrate_orig}
\lambda_k = \frac{1}{\hat{R}_k}\frac{\text{d} \hat{R}_k}{\text{d} t}=
\frac{1+\bar{R}^{2k}}{1-\bar{R}^{2k}}k
\left(\frac{\mathcal{A}^2 b^2(1-k^2)}{12Ca\bar{R}^3}-
\frac{1}{b}\left(\frac{1}{2}\frac{\text{d}
  b}{\text{d}
  t}-\frac{1}{\bar{R}^2}\right)\right)-
\frac{1}{b}\left(\frac{1}{2}\frac{\text{d}
  b}{\text{d} t}+\frac{1}{\bar{R}^2}\right).
\end{equation}
For bubbles that are much smaller than the outer radius
of the cell, we have $(1+\bar{R}^{2k})/(1-\bar{R}^{2k}) \approx 1$,
allowing us to approximate the growth-rate as
\begin{equation}
\label{growthrate}
\lambda_k = 
\frac{k-1}{\bar{R}^2}
\left(
\frac{1}{b} - 
\frac{\mathcal{A}^2 b^2 k (k+1)}{12Ca\bar{R}}-
\frac{\bar{R}^2}{2b}\frac{\text{d} b}{\text{d} t}
\frac{k+1}{k-1} \right).
\end{equation}
In the absence of lifting, i.e. $b=1$ and $\text{d} b/\text{d} t=0$, 
this reduces to Paterson's classical expression for the growth rate in  
an infinitely large Hele-Shaw cell with constant 
gap width~\cite{Paterson81}. A positive 
rate of plate separation, $\text{d} b/\text{d} t > 0$, can be 
seen to reduce the growth rate of the perturbations and therefore 
stabilises the axisymmetric state.

We note that, in general, the instantaneous growth rate $\lambda_k$
of the small-amplitude perturbations varies with the evolving mean 
radius of the bubble. As a result, the range of
unstable wavenumbers (i.e. 
wavenumbers for which $\lambda_k > 0$) and the most unstable 
wavenumber (for which $d\lambda_k/dk = 0$) generally change throughout
the system's evolution. One of the key observations made by \citet{Zheng15} 
is that if the gap width has a power-law behaviour
of the form $b(t) \sim t^{1/7}$ then the mode with
the largest positive growth rate predicted by Paterson's analysis -- which is 
used somewhat inconsistently because its derivation assumes that 
$\text{d} b/\text{d} t = 0$ -- remains constant. 

To assess to what extent \citeauthor{Zheng15}'s conclusions are
affected by this inconsistency we note that for an arbitrary 
time-varying gap width $b(t)$, equations (\ref{axisym_radius}) 
and (\ref{growthrate}) show that perturbations with wavenumber 
$k$ decay if
\begin{equation}
\label{inequality}
\frac{\text{d} b}{\text{d} t} -
\frac{(k-1)}{(k+1)}\frac{b}{t}+\frac{\mathcal{A}^2 k (k-1)
}{12\sqrt{2}Ca}\frac{b^{9/2}}{t^{3/2}}>0.
\end{equation}
For $b(t) = b_1 \ t^{1/7}$ this condition becomes
\begin{equation}
\label{inequality2}
S_k = \frac{1}{7} - \frac{k-1}{k+1} + \frac{ k (k-1)}{J} > 0,
\end{equation}
where
\begin{equation}
\label{define_J}
J = \frac{12 \sqrt{2} Ca}{\mathcal{A}^2 b_1^{7/2}}
  = \frac{6 \mu Q^{3/2}}{\pi^{3/2}\gamma b_1^{*\, 7/2}}
\end{equation}
is precisely the control parameter appearing in \citeauthor{Zheng15}'s
analysis. The inclusion of the $\text{d} b/\text{d} t$ term in our analysis
results in the appearance of an additional term in the stability
criterion (\ref{inequality2}) (the constant $1/7$
on the left-hand-side of this equation)
but its presence does not re-introduce a time-dependence into
this condition. Thus \citeauthor{Zheng15}'s key observation
remains unchanged. Furthermore, the condition for a perturbation with 
wavenumber $k$ to have
a negative instantaneous growth rate (implying linear stability with respect
to such perturbations) can be expressed in terms of $J$ as
\begin{equation}
\label{J_cr_of_k}
J < J_{\rm cr}(k) = \frac{7 k (k^2-1)}{6k-8},
\end{equation}
while, for a given, sufficiently large value of $J$, the wavenumber
$k_{\rm max}$ of the perturbation with the largest positive growth 
rate satisfies
\begin{equation}
\label{our_k_max}
J = \frac{7}{6} \left( 3 k_{\rm max}^2-1 \right).
\end{equation}
These results again only differ slightly from those  
obtained by \citeauthor{Zheng15}'s approach which yields $J_{cr}(k) = k(k+1)$
and $J = 3 k_{\rm max}^2-1$.

\begin{figure} 
 \includegraphics{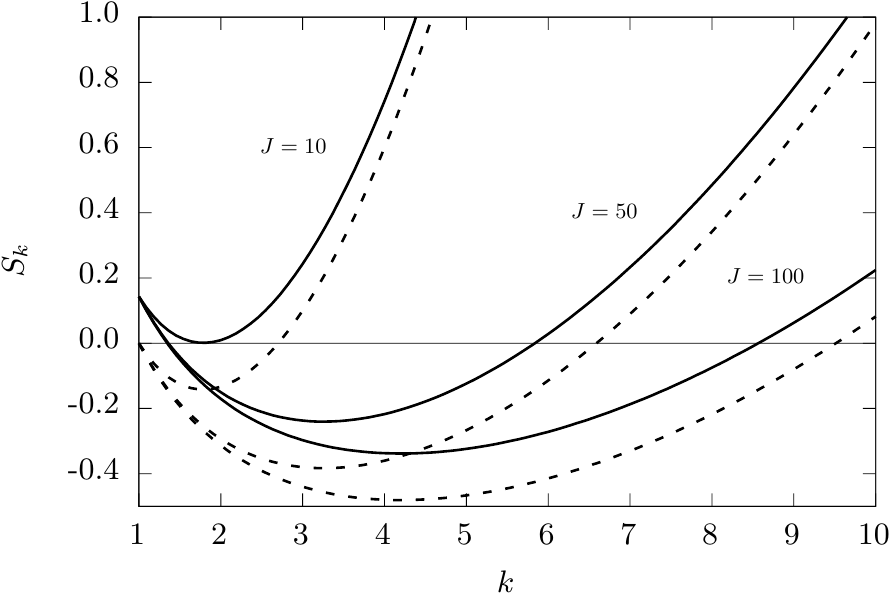}
\caption{\label{fig:bla}
Plot of the stability criterion (\ref{inequality2}) for
$J=10, 50$ and $100$. The axisymmetric state is unstable to
non-axisymmetric perturbations with integer wavenumber $k$ if $S_k < 0$.
The predictions corresponding to \citeauthor{Zheng15}'s 
approach \cite{Zheng15} are shown by dashed lines.
}
\end{figure}

Figure \ref{fig:bla} compares the two sets of results in a plot of the 
stability criterion (\ref{inequality2}) for three different values of
$J$. Recall that the axisymmetric state is stable when $S_k > 0$,
therefore wavenumbers for which $S_k < 0$ represent unstable
perturbations.  \citeauthor{Zheng15}'s approach can be seen to 
consistently over-estimate the growth-rate of the non-axisymmetric 
perturbations because their analysis omits the stabilising effect
of the $\text{d} b/\text{d} t$ term. However, both analyses show
that the range of unstable wavenumbers and the wavenumber with the 
fastest growth rate increase with $J$.

\section{\label{DNS}Direct numerical simulations}
The results presented above confirm that for a lifting law of the form
proposed by \citet{Zheng15}, the wavenumber of the
most rapidly growing small-amplitude perturbation remains constant 
throughout the system's evolution.  To assess if
this behaviour is responsible for the development of the non-splitting
fingers observed in \citeauthor{Zheng15}'s experiments, we now conduct
numerical simulations of the system's nonlinear evolution following
the onset of the linear instability.  For this purpose we used an {\tt
  oomph-lib}-based \cite{HeilHazelOomph2006} finite-element
discretization of the governing equations, details of which can be
found in \cite{PihlerPuzovic13}. 

All simulations were performed for 
$Ca = 0.3893$ and $R_{\rm init} = 0.05$
as in our previous work \cite{PihlerPuzovic13, PihlerPuzovic14}. 
Suitable temporal and spatial convergence tests were performed to 
ascertain that the results presented below are fully converged.

\subsection{Unimodal perturbations and self-similar fingering}
\begin{figure}
  \centering
   \includegraphics{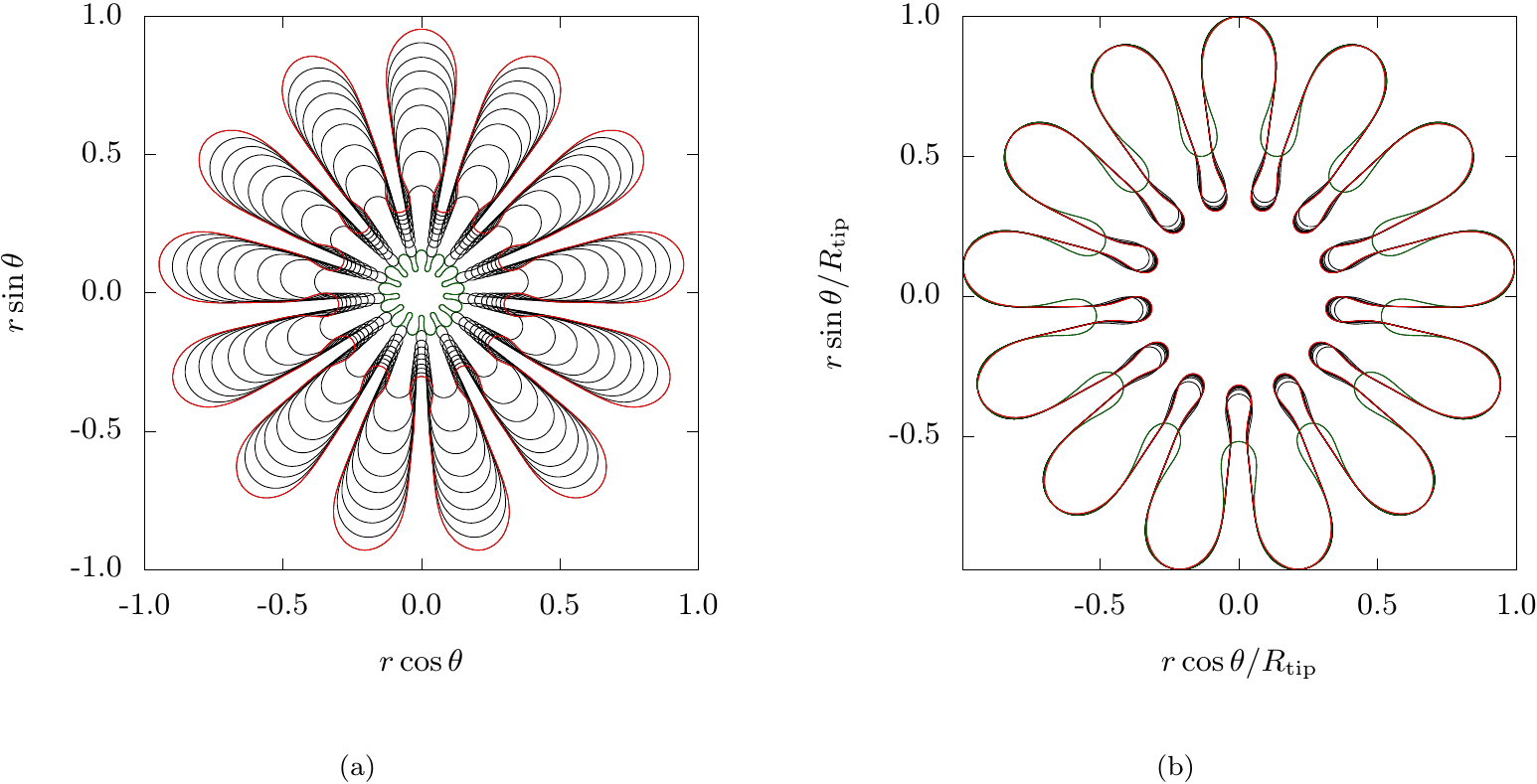}
  \caption{(a) Series of snapshots showing the evolution of the
    interface for $J=600$. The simulation was started at $t_0=
    0.00125$ when the axisymmetric interface (of radius $R_{\rm init}
    = 0.05$) was subjected to an initial perturbation 
    with wavenumber $k = k_{\rm max} = 13$ and amplitude $\epsilon=5 \times
    10^{-4}$. The innermost
    interface is shown for $t = 0.011$ and the time interval between
    contours is $\Delta t = 0.083$. (b) Interface contours from (a)
    scaled by the radius of the finger tip, $R_{\rm
    tip}$. In (a) and (b), the first and the last interfaces in 
    the sequence are highlighted with green and red colours, 
    respectively.\label{fig:interface_contours_J=600} }
  \end{figure}

We start by perturbing the initial circular bubble
of radius $R_{\rm init} = 0.05$ with the single, most
rapidly growing mode so that
\begin{equation}\label{per_single}
R (\theta, t=t_0)= R_{\rm init} + \epsilon \cos(k_{\rm max}\theta),
\end{equation}  
where we set the amplitude of the perturbation to 
$\epsilon=5 \times 10^{-4}$, i.e. 1\% of the
initial radius. Figure~\ref{fig:interface_contours_J=600}(a) shows the
time-evolution of the interface for $J=600$ (for which $k_{\rm max}=13$),
using snapshots of the interface plotted at regular time intervals. 
Thirteen identical finite-amplitude fingers grow from the initial perturbation
and, interestingly, the fingers show no signs of splitting. Instead, they remain
symmetric about their radial centreline and appear to approach a 
self-similar shape. This is confirmed in figure~
\ref{fig:interface_contours_J=600}(b), where we rescaled the radial coordinate
by the radius of the finger tip,
$R_{\rm tip}$, for each snapshot. This shows  that, after an initial transient
evolution that occurs within the first five contours shown in
figure~\ref{fig:interface_contours_J=600}(b), the successive rescaled interfaces
become virtually indistinguishable. Equivalent behaviour was
observed in simulations for 14 different values of $J$ in the 
range $25 \le J \le 1000$, where in each case we
imposed the most rapidly growing single-mode perturbation, with 
$k_{\rm max}$ given by equation (\ref{our_k_max}).

\begin{figure}
   \includegraphics{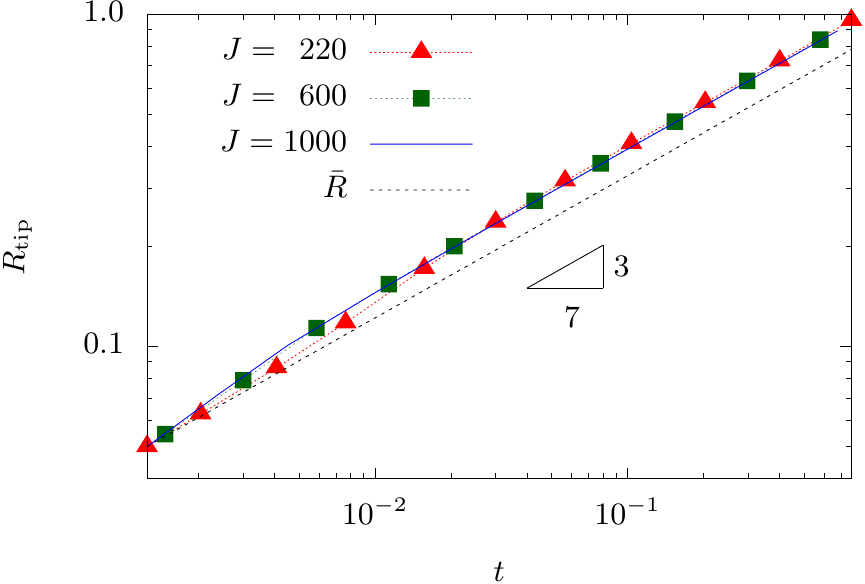}
  \caption{Log-log plot of $R_{\rm tip}$ as a function of time for
    $J=220$, 600 and 1000. In each case, the interface is initially perturbed by
    the single, most-rapidly growing mode. The dotted line shows the radius of 
    the axisymmetrically growing bubble, $\bar{R}(t)$.}
  \label{fig:tip_position_vs_time}
\end{figure}

In figure \ref{fig:tip_position_vs_time} we plot the tip radius,
$R_{\rm tip}$, as a function of time on a log-log scale for
a range of $J$ values. As the tip radius increases from its
initial value it rapidly approaches a power law behaviour,
$R_{\rm tip}\sim t^{3/7}$, which is identical to that of the
axisymmetrically growing bubble (see equation (\ref{axisym_radius})
and recall that $b(t) \sim t^{1/7}$). This is again consistent
with the observed self-similar evolution of the interface shape -- 
if the bubble radius $R(t,\theta)$ approaches a self-similar
behaviour such that $R(t,\theta) = f(t) F_k(\theta)$ volume
conservation requires that $f(t) \sim  t^{3/7}$. We note that the
curves representing $R_{\rm tip}(t)$ for 
different values of $J$ almost overlap, implying that
the pre-factor (which is always larger than that for the axisymmetrically
growing bubble) is approximately independent of $J$, despite significant
variations in $k_{\rm max}$ which ranges from 8 to 17.
   
\begin{figure}[h!]
   \includegraphics{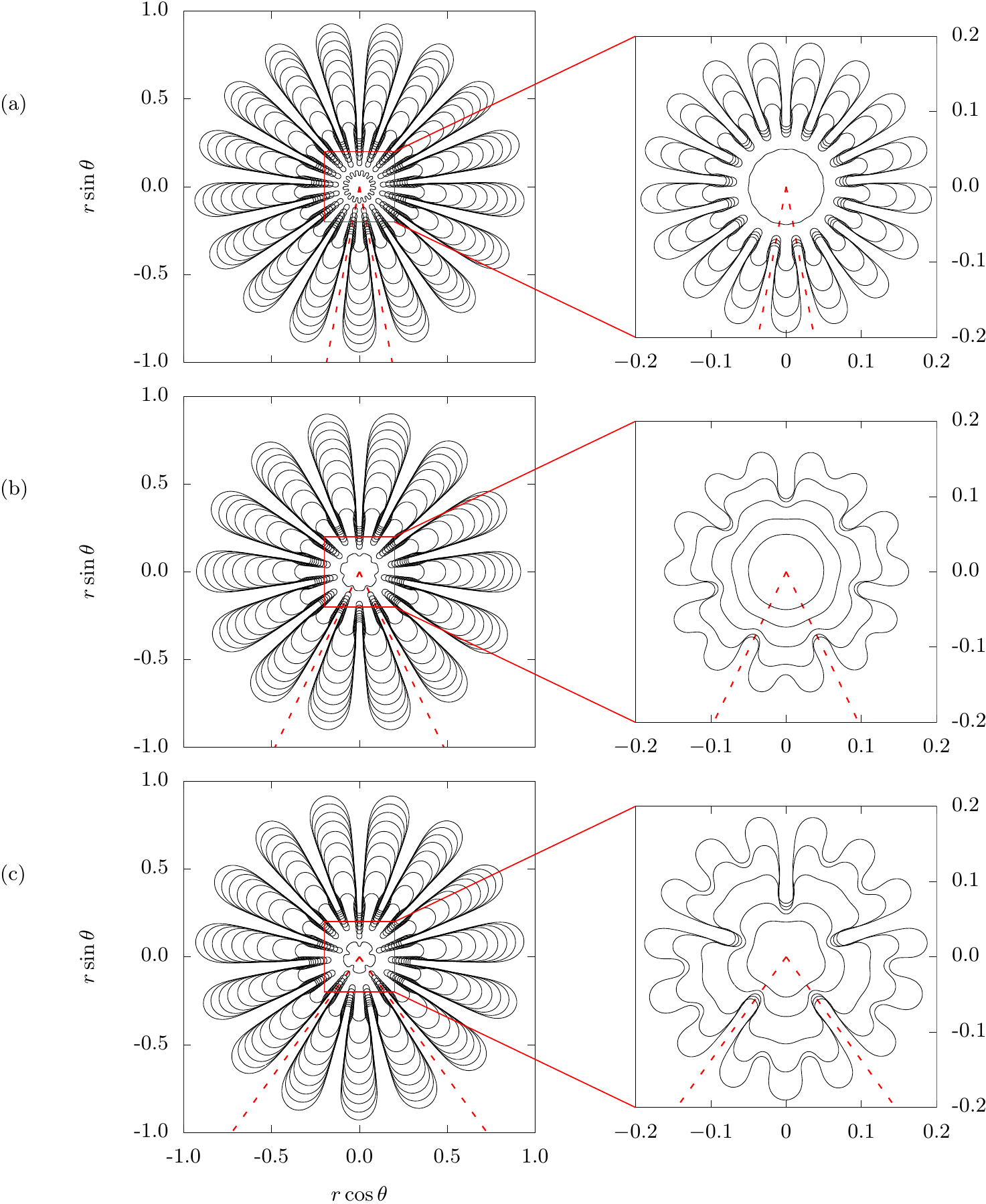}
  \caption{Series of snapshots showing the
    evolution of the interface
    for $J=1000$. The simulations were started at $t_0=
    0.00125$ when the axisymmetric interface (of radius $R_{\rm init}
    = 0.05$) was subjected to an initial perturbation 
    with wavenumber
    (a) $k = k_{\rm max} = 17$, (b) $k = 7$, (c) $k = 5$. The 
    amplitude of the initial perturbation is $\epsilon=5 \times
    10^{-4}$ for (a-b) and $1.25 \times 10^{-2}$ for (c). The time
    interval between contours is (a) $\Delta t = 0.069$, 
    (b) $\Delta t=0.078$, (c) $\Delta t=0.08$. 
    In each case the first contour is shown at $t=0.011$. Dashed lines 
    indicate a sector within which a single finger develops at early
    times. The insets show the early evolution of the
    interface with successive contours chosen so that the average
    radius $\langle R\rangle =
    \frac{1}{2\pi}\int_0^{2\pi}R\text{d}\theta$ increases by 0.022.}
  \label{fig:interface_contours_J=1000}
  \end{figure} 

Next, we explore whether the interface shape evolves towards a
self-similar solution when the initial 
single-mode perturbation has a wavenumber other than $k_{\rm max}$. 
For this purpose we choose a larger value of $J$,
$J=1000$, in order to operate in a regime with a wide range of linearly unstable
modes. We then perturb the initial circular bubble as in
(\ref{per_single}) but replace $k_{\rm max}$ with a wavenumber from
the unstable range defined by equation (\ref{J_cr_of_k}). 
The resulting evolution of the interface is shown in
figures~\ref{fig:interface_contours_J=1000}(a-b) for different values of
$k$, starting from a perturbation of amplitude $\epsilon=5 \times 10^{-4}$.
In each case, the system undergoes a transient initial evolution
but ultimately reaches a regime in which a number
of identical, self-similar fingers have emerged from the initial
perturbation. However, the number of self-similar fingers that develop
depends on the wavenumber of the initial perturbation. For the perturbation 
with $k=k_{\rm max} =17$
(figure~\ref{fig:interface_contours_J=1000}(a))  
we retain 
the wavenumber of the initial perturbation. If we impose an initial 
perturbation with $k=7$ (figure~\ref{fig:interface_contours_J=1000}(b)) the 
finite-amplitude fingers that develop from the initial perturbation 
split once (see the zoomed-in region shown in the inset) before
approaching a self-similar pattern with 14 fingers.
This implies that the number of self-similar fingers that can 
emerge from the initial perturbation is not directly related to 
the value of $k_{\rm max}$ from the linear stability analysis.

The tip splitting events that may occur during the early transients 
that precede 
the self-similar evolution of the interface are not restricted to tip doubling. 
This is illustrated in figure~\ref{fig:interface_contours_J=1000}(c) where
we imposed an initial perturbation with $k=5$ and also increased 
the amplitude of the perturbation to 25\% of the initial 
radius. The early-time growth of the initial perturbation is now
followed by a tip-tripling event (see the zoomed-in region shown in
the inset) before the bubble approaches a 
configuration with 15 self-similar fingers.

For all cases shown in figure~\ref{fig:interface_contours_J=1000} 
the imposed unimodal initial perturbation is symmetric about the 
radial centreline of a sector spanned by an angle $2\pi/k$, shown by
the dashed lines. Even when the evolution of the interface 
involved tip-splitting into two or three fingers, the newly-formed 
fingers did not compete with each other to break this initial symmetry.
The fact that the computations were performed with unstructured 
meshes which (despite being sufficiently  fine to ensure
mesh-independent solutions) inevitably introduce small non-axisymmetric
perturbations, suggests that the self-similar fingers are linearly
stable.

\subsection{Non-unimodal perturbations and disordered fingering}
Unimodal initial perturbations, such as those imposed in the simulations 
presented in the previous section, are, of course, difficult to realise 
in actual experiments. The initial perturbation to a real bubble are 
likely to comprise a wide range of modes. To analyse this scenario 
we now perturb the interface with the full range of
linearly-unstable wavenumbers. We introduce these
perturbations by adding all of the unstable modes to the initial shape
of the interface, each with a randomly generated (and uniformly
distributed) relative amplitude $\hat{R}_k \in [0 , 1]$ and phase $\phi_k \in
[0 , 2\pi)$,
\begin{equation}
R(\theta, t=t_0) = R_{\rm init} + \epsilon \sum_{\mathrm{unstable} \ k}
 \hat{R}_k\cos(k\theta + \phi_k),
\end{equation} 
where the overall amplitude $\epsilon$ was chosen to be 0.5\% of the
initial radius. We performed 50 simulations for 14 values of $J$
within the range $25 \le J \le 1000$.
 
\begin{figure}
  \centering
  \includegraphics{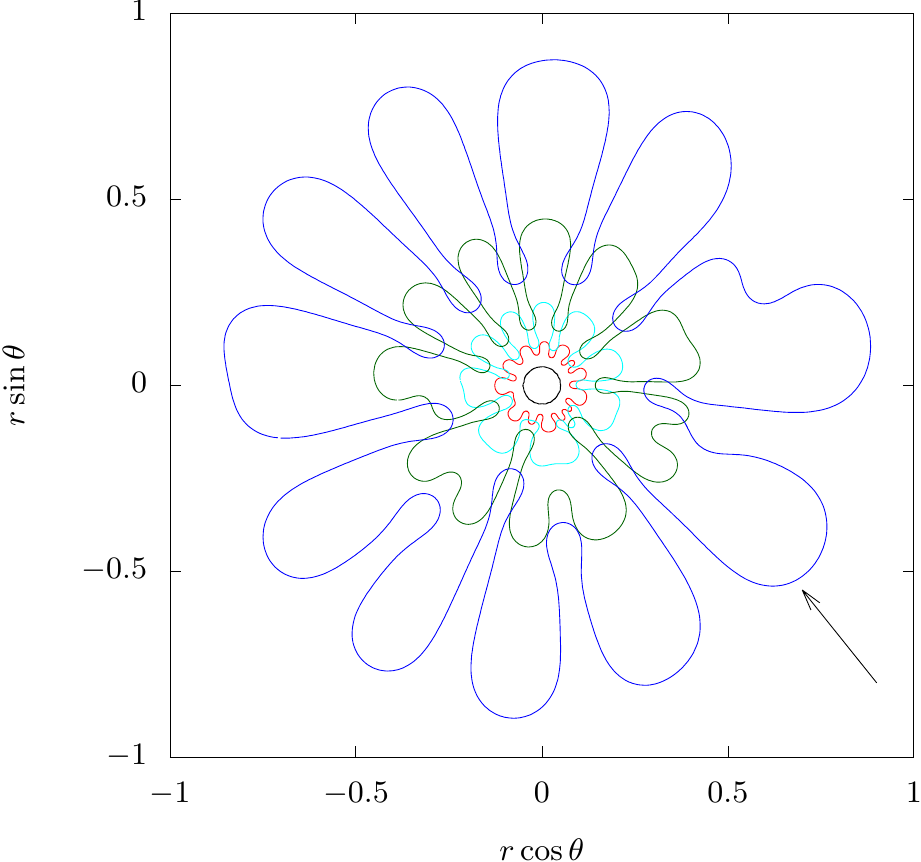}
\caption{Instantaneous interface shapes with average radius $\langle
  R \rangle=0.05$ (black), 0.1 (red), 0.17 (cyan), 0.31 (dark green)
  and 0.6 (blue), respectively, computed for $J=600$.  The arrow
  points to a region of the interface that undergoes tip-splitting and
  finger competition.\label{fig:contours_and_ffts}}
\end{figure}
 
\begin{figure}[h!]
  \centering
  \includegraphics{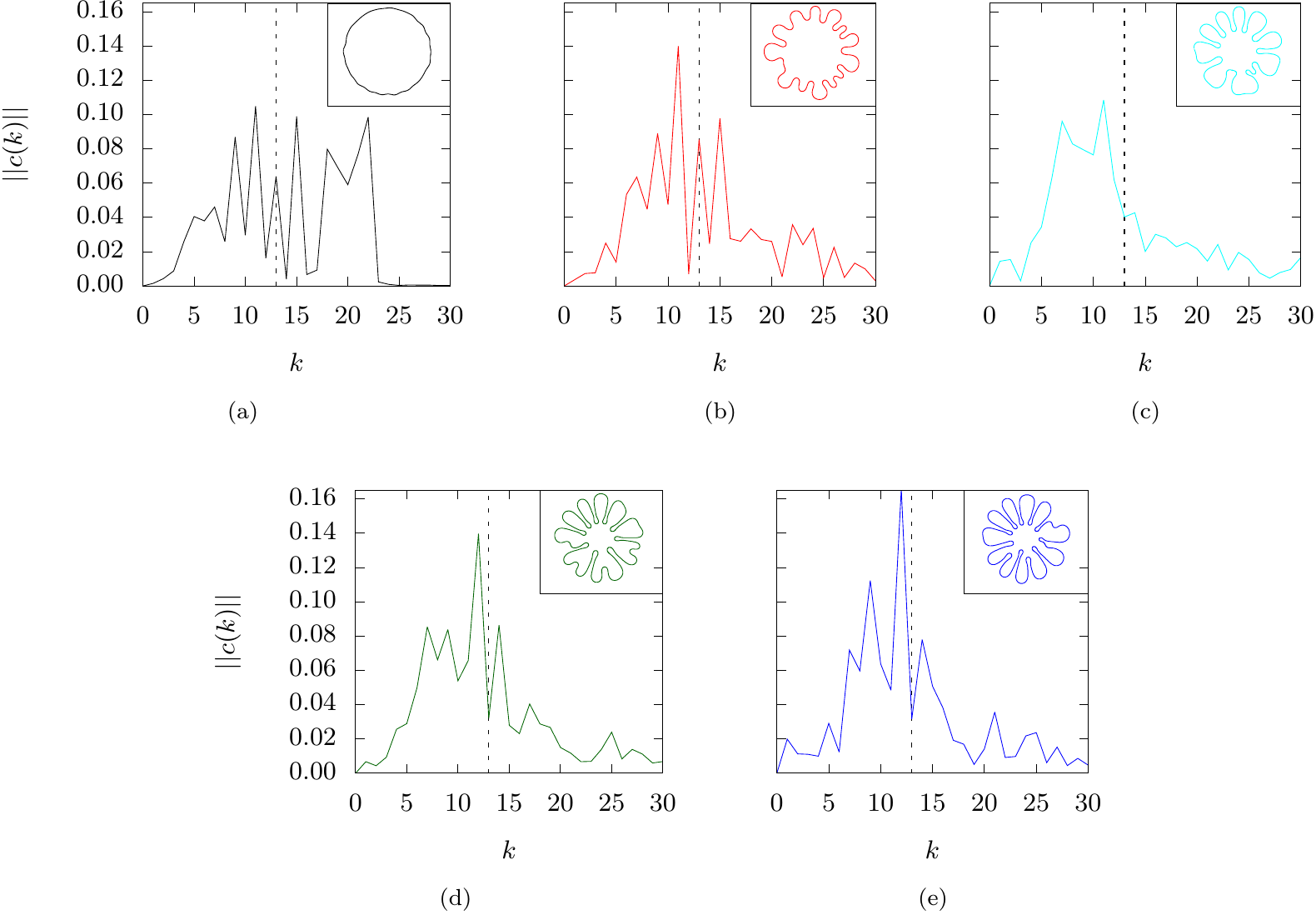}
\caption{Normalised spectra of the instantaneous interface
  shapes shown in figure~\ref{fig:contours_and_ffts} for $J=600$. The
  mean radii are (a) $\langle R \rangle=0.05$, (b) 0.1,
  (c) 0.17, (d) 0.31 and (e) 0.6. The vertical dashed line indicates
  the value of $k_{\rm max} = 13$ predicted by the linear stability analysis.}
\label{fig:ffts}
\end{figure}
 
A typical evolving interface is shown in figure
\ref{fig:contours_and_ffts} for $J=600$ for which $k_{\rm max} = 13$. The 
snapshots of the interface are shown for different values of the average 
radius $\langle R\rangle$ and are plotted in different colours. 
When the mean radius has grown to $\langle R\rangle = 0.1$ (red) the
random initial perturbation has led to the development of 13 distinct 
fingers of varying widths and lengths. Differences in the length 
of adjacent fingers are amplified as the fingers grow (finger
competition) and when $\langle R\rangle =
0.17$ (cyan), the total number of fingers has decreased from
13 to 10.  At $\langle R\rangle = 0.17$ (cyan) the four largest fingers have
broadened considerably so that their tips are approximately flat. When
$\langle R\rangle = 0.31$ (dark green) these fingers have
undergone a tip-doubling followed by competition between the 
newly-formed fingers, while another finger has broadened sufficiently for
another tip-doubling to occur. Because the fingers are not symmetric
about their radial centreline, the tip-doubling events lead to pairs
of fingers of different shapes and radial tip positions, so that one
of the fingers screens the other and adopts a strongly asymmetric
shape; see, for example, the finger identified by the arrow in figure 
\ref{fig:contours_and_ffts}. The
final pattern shown in figure \ref{fig:contours_and_ffts} (blue) at $\langle
R\rangle = 0.6$ comprises 11 distinct fingers of different shapes, which
continue to evolve and interact as they grow.

We characterise the patterns in figure \ref{fig:contours_and_ffts} 
by following \cite{Zheng15} and compute the Fourier coefficients 
$c(k)$ of $R - \langle R\rangle$,
\begin{equation}\label{Ft}
c(k) = \frac{1}{2\pi} \int_{-\pi}^{\pi}(R - \langle R\rangle)\exp(-ik\theta) \text{d}\theta.
\end{equation}
To compare coefficients computed for different values of $\langle
R\rangle$, we normalise the amplitudes $c(k)$ by the area under the
corresponding spectral curves and denote the normalised amplitudes by
$||c(k)||$. This allows us to focus on the relative amplitudes of the
modes rather than their absolute value which grows with increasing
radius.

Figure \ref{fig:ffts} shows that the evolution of the spectral curves is
closely correlated with the evolution of the interfacial pattern. 
The interface shape can be seen to comprise a mixture of modes, with 
the number of distinct fingers observed at the interface approximately 
corresponding to the mode with the maximum normalised amplitude, 
which we label $K_{\rm max}$.
 
\begin{figure}[h!]
  \includegraphics{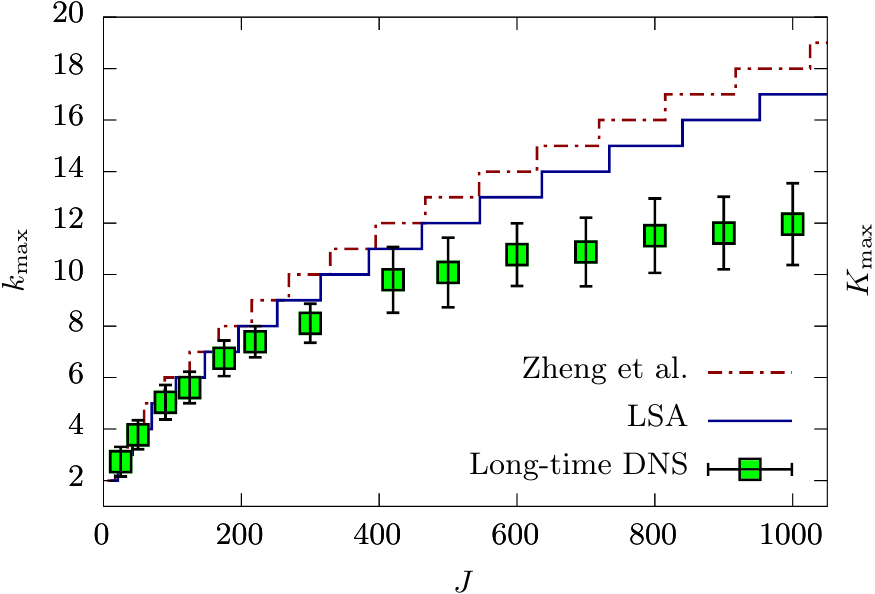}
  \caption{\label{fig:k-max_vs_J}
    The variation of $K_{\rm max}$ with $J$ from direct
    numerical simulations (DNS). The error bars denote the standard
    deviation across 50 simulations.
    The lines show $k_{\rm max}$ from the results of the linear
    stability analysis, using our own predictions (solid) and those from
    \cite{Zheng15} (dashed). $k_{\rm max}$ can only take integer
    values, which we connect by vertical line segments, resulting in
    steps in the curves. }
\end{figure}

In all the simulations performed with random initial perturbations the
value of $K_{\rm max}$ continued to fluctuate throughout the system's
evolution. Typically, $K_{\rm max}$ decreased significantly from its
initial value at early times when many modes had comparable amplitudes (see
figure~\ref{fig:ffts}(a)). At later
times, $K_{\rm max}$ was more likely to increase due to tip-splitting
events. The overall rate of change of $K_{\rm max}$ decreased with
increasing time, but $K_{\rm max}$ continued to
fluctuate until fingers reached the edge of the computational
domain. This scenario was observed for all the values of $J$
investigated ($J \in [25,1000]$), despite the fact that the growth
rate of the imposed small-amplitude initial perturbation is smaller
for smaller values of $J$.

Figure \ref{fig:k-max_vs_J} shows how variations in $J$ 
affect $K_{\rm max}$ which we evaluated when $\langle R\rangle=0.6$ 
-- the largest value reached in all simulations (which necessarily 
terminate when one of the fingers reaches the outer edge of the 
computational domain). The error bars on the symbols quantify the 
standard deviation of $K_{\rm max}$ across 50 runs for each value 
of $J$, and highlight the disordered nature of the interface evolution. 
The two lines show the most unstable wavenumber $k_{\rm max}$ 
predicted by our linear stability analysis (solid)
and by \citeauthor{Zheng15}'s approach (dashed). For modest values of
$J$ the wavenumber $K_{\rm max}$ 
observed in the numerical simulations is remarkably 
close to $k_{\rm max}$ from the linear stability analysis. 
However, for larger values of $J$, $K_{\rm max}$ 
remains significantly below $k_{\rm max}$. Furthermore, none of the 
simulations that were started with random initial perturbations 
showed any signs of settling on a self-similar state -- the 
fingers continued to split and then compete with each other over 
the entire range of the simulations.

Our results show that the system's behaviour is fundamentally
different when the initial, axisymmetrically expanding bubble
is subjected to unimodal or non-unimodal perturbations to its
shape. This raises the question in what way the 
small-amplitude non-unimodal perturbations that we deliberately 
introduced in the computations shown in figure \ref{fig:contours_and_ffts}
differ from those introduced unintentionally by the use of unstructured
meshes in the simulations in figure \ref{fig:interface_contours_J=1000}, say.
Clearly, the deliberately introduced unimodal 
perturbation imposed by equation (\ref{per_single}) 
must have a larger initial amplitude than those 
caused by the unstructured mesh. This was ensured by performing the
computations on sufficiently fine meshes. However, we also have to
make sure that the growth rate of the deliberately introduced unimodal 
perturbation is sufficiently large so that it deforms the interface 
into a non-linear regime before the other modes have grown to
comparable amplitude. This is, of course, easiest to achieve 
if the deliberately introduced perturbation is the one with the 
maximum growth rate, as in figure \ref{fig:interface_contours_J=1000}(a)
where we applied a perturbation with $k = k_{\rm max} = 17$.
The initial amplitude of $\epsilon = 5 \times 10^{-4}$ used for the 
perturbation with $k = 7 < k_{\rm max}$ in figure 
\ref{fig:interface_contours_J=1000}(b) sufficed to create
seven finite-amplitude fingers that subsequently underwent a single
tip-splitting event before approaching the self-similar regime. 
However, when using that amplitude to impose a perturbation with $k=5$
(which has a much smaller but still positive growth rate) the system
evolved in a disordered manner, with unequal fingers that kept 
splitting and competing with each
other. The five-fold overall symmetry of the initial perturbation 
could only be retained by increasing its amplitude, as in figure
\ref{fig:interface_contours_J=1000}(c) where the initial fingers
undergo a tip-tripling before approaching a self-similar regime
with 15 non-splitting fingers.

\section{\label{conclusions}Summary and discussion}
We studied the dynamics of an expanding gas bubble
that displaces a viscous liquid in a radial Hele-Shaw cell with a
time-varying gap width, focusing on the case of a $t^{1/7}$ 
power-law for the plate separation. Our full linear stability analysis 
confirmed \citeauthor{Zheng15}'s \cite{Zheng15} prediction that the
wavenumber $k_{\rm max}$ of the most unstable small-amplitude perturbation to 
the axisymmetrically expanding bubble is independent of the bubble's
radius. We then employed numerical simulations to follow the growth of the 
instability into the finite-amplitude regime. This showed that for
unimodal perturbations with a wavenumber that matches the most
unstable wavenumber from the linear stability analysis,
self-similar fingers formed on the interface. This is in stark contrast
to the typical behaviour observed in Hele-Shaw cells with constant gap widths
where fingers tend to split and compete with each other, resulting
in complex, disordered and continuously evolving interface shapes.

Interestingly, self-similar fingers also emerged from
unimodal perturbations with wavenumbers other than $k_{\rm max}$. In
this case, the finite-amplitude fingers tended to pass through an initial
transient period during which they split (via tip doubling or
tripling) before ultimately approaching a final regime in which their
shapes evolved in a self-similar fashion without any further 
splitting. The number of identical finite-amplitude, self-similar 
fingers emerging from this process depended on the initial condition. 
The self-similar fingers appear to be linearly stable in the sense that they
persist despite inevitable small perturbations due to the use of
unstructured meshes in the simulations. 
The fact that the self-similar solutions can only be realised from unimodal 
perturbations with wavenumber $k$, so that the interface was 
symmetric about the radial centreline of a sector
spanned by an angle $2\pi/k$,  suggests
that they are unlikely to be observable in actual experiments 
(unless an experiment happens to be performed in the narrow regime 
where there is only a single unstable mode). 

When starting the simulations from non-unimodal, random 
perturbations to the initial bubble shape -- a scenario more 
representative of the situation in actual physical experiments --
multiple, competing and continuously splitting fingers developed,
similar to the behaviour typically found in the constant-gap system. 
For modest values of the control parameter $J$, the wavenumber 
$K_{\rm max}$ of the dominant mode contained in the Fourier spectrum
of the finite-amplitude fingers tended to be close to $k_{\rm max}$ (as
observed in \citeauthor{Zheng15}'s experiments) but for larger values
of $J$ we found $K_{\rm max}$ to be significantly smaller than $k_{\rm
max}$, suggesting that there is no straightforward relationship
between the two quantities. In fact, it is unclear how the fact that
the wavenumber of the most-unstable small-amplitude perturbation to 
the axisymmetric bubble remains independent of its radius could 
affect the behaviour of the finite-amplitude fingers that emerge from the 
linear instability: tip-splitting of finite-amplitude fingers occurs in the
course of their non-linear evolution (and does not necessarily involve
the occurrence of a secondary instability), rather than because the
most unstable wavenumber of perturbations to some (far away)
axisymmetric state changes. We therefore suggest to interpret 
the non-splitting fingers observed in this system as self-similar 
solutions of the governing equations. They are analogues to the 
steadily propagating fingers that develop in rectangular Hele-Shaw channels
\cite{Saffman58, BenAmar91a, BenAmar91b, andres}.
We observed the systematic transient evolution of the interface 
towards these self-similar states via tip-doubling and
tripling. By analogy with rectangular Hele-Shaw channels, this 
suggests that in radial Hele-Shaw cells other unstable self-similar 
solutions may exist which are reminiscent of the unstable 
Romero--Vanden-Broeck multiple-tip solutions~\cite{andres}.

\begin{acknowledgments}
The authors would like to thank Andrew Hazel, Scott McCue and Liam
Morrow for helpful discussions. The work of A.J. was funded through EPSRC grant
EP/P026044/1. The research
data supporting this publication can be found in the supplementary material.
\end{acknowledgments}
\bibliography{references}
\end{document}